\documentclass[aip,jcp,amsmath,amssymb,reprint,numeric]{revtex4-1}
\usepackage{graphicx}
\usepackage{subfigure}
\usepackage{float}
\usepackage{amsmath}

\begin{document}
\title[]{Continuous time random walk concepts applied to extended mode coupling theory: A study of the Stokes-Einstein breakdown. }

\author{Manoj Kumar Nandi}
\address{\textit{Polymer Science and Engineering Division, CSIR-National Chemical Laboratory, Pune-411008, India}}
\author{Sarika Maitra Bhattacharyya}
\email{mb.sarika@ncl.res.in}
\address{\textit{Polymer Science and Engineering Division, CSIR-National Chemical Laboratory, Pune-411008, India}}



\begin{abstract}
In an attempt to extend the mode coupling theory (MCT) to lower temperatures, an Unified theory was proposed which within the MCT framework incorporated the activated dynamics via the random first order transition theory (RFOT). Here we show that the theory although successful in describing other properties of supercooled liquids is unable to capture the Stokes-Einstein breakdown. We then show using continuous time random work (CTRW) formalism that the Unified theory is equivalent to a CTRW dynamics in presence of two waiting time distributions. It is known from earlier work on CTRW that in such cases the total dynamics is dominated by the fast motion. This explains the failure of the Unified theory in predicting the SE breakdown as both the structural relaxation and the diffusion process are described by the comparatively fast MCT like dynamics. The study also predicts that other forms of extended MCT will face a similar issue. We next modify the Unified theory by applying the concept of renewal theory, usually used in CTRW models where the distribution has a long tail. According to this theory the first jump given by the persistent time is slower than the subsequent jumps given by the exchange time. We first show that for systems with two waiting time distributions even when both the distributions are exponential the persistent time is larger than the exchange time. We also identify the persistent time with the slower activated process. The extended Unified theory can now explain the SE breakdown. In this extended theory at low temperatures the structural relaxation is described by the activated dynamics whereas the diffusion is primarily determined by the MCT like dynamics leading to a decoupling between them. We also calculate a dynamic lengthscale from the wavenumber dependence of the relaxation time. We find that this dynamic length scale grows faster than the static length scale.
\end{abstract}

\maketitle

\section{Introduction}

When a liquid is cooled below its melting point it enters a supercooled liquid regime. Although the state of the system still remains in a liquid form, the characteristics of the supercooled liquid becomes quite different from that of a high temperature liquid. One of the characteristics that changes is the validity of the Stokes-Einstein (SE) relationship \cite{hansen_mcdonald,einstein,landau_fluid_mech}. At high temperatures the diffusion coefficient and the shear viscosity/structural relaxation of a liquid follow each other and this is expressed via the SE relation. However in the supercooled regime, the dynamics of the system changes and it is no more only diffusive in nature. As the temperature of the system is lowered the underline landscape properties start influencing the dynamics \cite{srikanth,Srikanth_nature}. The presence of the free energy barriers in the landscape are now experienced by the system. Thus according to the landscape picture the dynamics in a basin is described by diffusion and  the dynamics in between two basins is activated in nature. As the temperature is lowered the barriers become higher and the activation process becomes slower. It also becomes the dominant contributor to the structural relaxation. 
Recent experiments \cite{schmidt1991nature,fujara1992translational,cicerone1995relaxation,williams1996determination} and computer simulation studies \cite{hurley1995kinetic,mel1995long,kob1997dynamical} have also shown that unlike in normal high temperature liquids the dynamics of supercooled liquids is not homogeneous. In a given interval of
time, there are `mobile' and `immobile' regions giving rise to dynamical heterogeneity. The particles in these regions move 
cooperatively and the regions are called cooperatively rearranging region (CRR).
The slowing down of the dynamics in a supercooled liquid is also associated with a growing length scale of these CRRs. According to random first order transition theory (RFOT) the growth of this lengthscale and the increase in the free energy barrier are correlated \cite{lubchenko2003barrier}.  It is believed that due to this change in the dynamics a decoupling between the diffusion and the viscosity/structural relaxation takes place and  often the SE breakdown is considered a hallmark of a supercooled liquid \cite{SE_shiladitya,SE_hocky,SE_flenner,SE_staley}.

Because of the change in the dynamics the microscopic theories which work well in the normal liquid regime fails in the supercooled domain. 
It is know that amongst the existing microscopic theories mode coupling theory
is one of the best to describe the dynamics in moderately supercooled regime \cite{gotze1999}. However, it fails to describe the low temperature dynamics of the system specially below the dynamic transition temperature \cite{szamel-pre}. The absence of the activated dynamics in the theory is usually connected to its failure.  
The RFOT on the other hand can describe the activated dynamics in this temperature regime \cite{lubchenko2003barrier}. However, as mentioned above, both diffusive and activated motions contribute to the dynamics at low temperatures. There is no single theory which can microscopically describe both these motions. Few attempts have been made in this direction by extending the MCT and incorporating the activated dynamics \cite{gotze1,gotze2,bridging,chong_pre}.
One of them is the Unified theory\cite{bridging,sarika_PNAS}. Unlike the other theories \cite{gotze1,gotze1,chong_pre} the Unified theory connects the dynamics at low temperatures to the thermodynamics of the system. The activated dynamics in this theory is calculated using concepts of RFOT \cite{lubchenko2003barrier} which predicts that the free energy barrier of activation depends on the configurational entropy. This dependence of the activated dynamics on entropy is similar to what has been predicted by the Adam Gibbs theory \cite{adam-gibbs} and have also been observed in many simulation studies \cite{shila-jcp,atreyee_prl}. The unified theory has been successful in extending the dynamics described by MCT to much lower temperatures where the idealized MCT shows no relaxation \cite{bridging,sarika_PNAS}. It has been shown that the modified MCT coupled to the activated dynamics predicts hopping induced diffusion. Given the success of the theory it is imperative to test its validity in explaining the SE breakdown in supercooled liquids. As discussed before, there has been other attempts to extent the MCT \cite{gotze1,gotze2,chong_pre}. In one such study the breakdown of the SE was addressed \cite{chong_pre}. Except for the way the activated dynamics is calculated this formalism is quite similar to the Unified theory. Although the formalism could predict the SE breakdown it is also mentioned that the predicted decoupling between the diffusion and the structural relaxation is quite weak and the theory fails to explain the strong decoupling observed in experiments \cite{ediger_jpcb}.

In this article we first address the SE breakdown using the existing form of the Unified theory. We find that similar to that observed by Chong\cite{chong_pre} the Unified theory shows a weak or in this case almost no decoupling between the relaxation time and the diffusion. A continuous time random walk (CTRW) analysis is performed to show that the Unified theory is like a CTRW dynamics where there are two different processes with different waiting time distributions. In CTRW formalism it is known that when two processes have widely different timescales the dynamics is dominated by the fast process \cite{ctrw_2dis,tachiya_bagchi_seki}. This explains why the Unified theory does not predict the SE breakdown as both the structural relaxation and the diffusive dynamics are described by the fast MCT like dynamics. Thus we extend the Unified theory by using the concepts of renewal process in CTRW \cite{renewal_theory}. Earlier this concept has been extensively used by kinetically constrained model (KCM)  \cite{Chandler}. The extended Unified theory does predict a strong breakdown of the SE relation. We show that at low temperatures the structural relaxation is determined by the activated motion whereas the diffusion is described by the MCT like dynamics. This decoupling in the two dynamics leads to the SE breakdown. The theory can also predict the dynamic correlation length of the CRR. We find that the dynamic correlation length grows faster than the static correlation length. Our study clearly shows that for all extended MCT forms it is important to include the concepts of renewal theory to describe the true dynamics of the system.

The organization of the rest of the paper is the following. In the next section we describe the theoretical developments. In Sec. III we present results and discussions and Sec. IV contains a brief summary.


\section{Theory}
\subsection{Extended MCT and structural relaxation}
 We start our analysis from the ideal mode coupling theory (IMCT). The equation of motion for the total intermediate scattering function $\phi_{MCT}^{id}(q,t)$ is given by,
\begin{eqnarray}
 \ddot\phi_{MCT}^{id}(q,t)+\gamma \dot\phi_{MCT}^{id}(q,t)+\Omega_q^2\phi_{MCT}^{id}(q,t)\nonumber\\
 +\Omega_q^2\int_0^t{dt'\mathcal{M}^{id}(q,t')\dot\phi_{MCT}^{id}(q,t-t')}=0 .
\label{id-mct}
\end{eqnarray}
Here $\Omega_q^2=q^2k_BT/mS(q)$, $k_B$ is the Boltzmann's constant and $\gamma$ is the binary friction term, which is the short time part 
of the memory kernel. The memory kernel
$\mathcal{M}^{id}(q,t)$ describes the coupling between dynamics at different wavenumbers and can also be visualized as the  correlation between fluctuating forces \cite{Kawasaki2003}. This can be written as,
\begin{eqnarray}
\mathcal{M}^{id}(q,t)=\frac{1}{2(2\pi)^2\rho q^2}&\int& {d\bf{k}}V_q^2({\bf{k,p}})S(q)S(p)S(k)\nonumber\\
                     &\times& \phi_{MCT}^{id}(k,t)\phi_{MCT}^{id}(p,t).
\label{id-mct-memory}                     
\end{eqnarray}
Where $\bf{p=q-k}$ and $V_q({\bf{k,p}})=[({\bf{q.k}})C(k)+({\bf{q.p}})C(p)]$. Here $S(q)$ is the static structure factor and the direct correlation function, 
$C(q)=(1-1/S(q))/\rho$. Introducing Laplace transform, defined as $\phi(q,s)=\mathcal{L}[\phi(q,t)]$, we can write Eq.\ref{id-mct}
in equivalent form as,
\begin{eqnarray}
 \phi_{MCT}^{id}(q,s)=\frac{1}{s+\frac{\Omega_q^2}{s+\eta_l^{id}(q,s)}}.
 \label{id-laplace}
\end{eqnarray}
Here the longitudinal viscosity $\eta_l^{id}$ is given by,
\begin{eqnarray}
 \eta_l^{id}(q,s)=\gamma +\Omega_q^2\mathcal{L}[\mathcal{M}^{id}(q,t)].
 \label{etal-id}
\end{eqnarray}
Eq.\ref{id-mct} and Eq.\ref{id-mct-memory} can be solved iteratively. Given the information of the structure of the liquid IMCT can describe its dynamics. The temperature where we find a dynamical arrest $T=T_c^{micro}$ is the glass transition temperature of the system as predicted by IMCT. Studies have shown that $T_c^{micro}$ is higher than $T_g$ \cite{szamel-pre,tarjus_pre,unravel,role_pair}. It is been conjectured that the origin of this failure of IMCT in predicting the correct transition temperature is due to the exclusion of
hopping motion \cite{bridging}.

For supercooled liquids hopping creates an extra relaxation channel. In an earlier work by one of us an attempt had been made to include the hopping motion within the IMCT \cite{bridging}. In this scheme, usually referred to as Unified (MCT+RFOT) theory
the total intermediate scattering function in supercooled regime can be written as,
\begin{equation}
  \phi(q,t)\simeq \phi_{MCT}(q,t)\phi_{hop}(q,t).
\label{phitot}
\end{equation}
Here $\phi_{MCT}(q,t)$ is the mode coupling part and $\phi_{hop}(q,t)$ is the contribution from the hopping motion. 

With the approximation that $\phi_{hop}(q,t)$ is exponential, $\phi_{hop}(q,t)=\exp(-K_{hop}t)$, equation of motion for $\phi(q,t)$ can be written as \cite{bridging},
\begin{eqnarray}
 &&\ddot\phi(q,t)+(\gamma+2K_{hop}) \dot\phi(q,t)+(K_{hop}^2+K_{hop}\gamma+\Omega_q^2)\nonumber\\
 &\times&\phi(q,t)+\Omega_q^2\int_0^t{dt'\phi_{hop}(q,t')\mathcal{M}(q,t')}\nonumber\\
 &\times&  {[\dot\phi(q,t-t')+K_{hop}\phi(q,t-t')]}=0.
\label{scheme-1}
\end{eqnarray}
\noindent
Where 
$\mathcal{M}(q,t)=\frac{1}{2(2\pi)^2\rho q^2}\int {d\bf{k}}V_q^2({\bf{k,p}}) 
S(k)S(p)S(q) \times \phi(k,t)\phi(p,t)$ and
$V_q({\bf{k,p}})=[({\bf{q.k}})C(k)+({\bf{q.p}})C(p)]$. Note that the memory 
function is dependent on full $\phi(q,t)$. The hopping term $K_{hop}$ is 
expressed as, $K_{hop}=\frac{v_0}{v_p}\frac{q^2l^2P_{hop}(\Delta F)}{(1+q^2l^2)S(q)}$ 
\cite{maam_q_depend}. $P(\Delta F)=\frac{1}{\tau_0}\exp(-\Delta F/k_BT)$, where $\Delta F$
is the Free energy barrier (Details of the calculation are given in Appendix II). 
Here $v_0$ is the volume of the correlated region with correlation length $l_{static}$. 
$l_{static}$ is the distance where $ \frac {\partial F}{\partial r}=0$.
$v_p$ is the volume of a single particle,  and $l$ is the most probable jump length
which is close to the Lindemann Length \cite{lindman}. Following the study of Chong
\cite{chong_pre} in the  expression of $K_{hop}$ we include the structure factor 
which was not present in the earlier study \cite{maam_q_depend}.

The equation of motion for $\phi_{MCT}(q,t)$, which is consistent with Eq.\ref{scheme-1} can be written as,
\begin{eqnarray}
 \ddot\phi_{MCT}(q,t)&+&\gamma \dot\phi_{MCT}(q,t)+\Omega_q^2\phi_{MCT}(q,t)\nonumber\\
                     &+&\Omega_q^2\int_0^t{dt'\mathcal{M}(q,t')\dot\phi_{MCT}(q,t-t')}=0.
\label{phi-prod}
\end{eqnarray}
Note that except for the memory function the above equation is same as the equation of motion for $\phi_{MCT}^{id}(q,t)$. The memory function $\mathcal{M}(q,t)$ is the same as in Eq.\ref{scheme-1}, dependent on the full $\phi(q,t)$. Thus the MCT dynamics is coupled to the hopping dynamics through the memory function. In earlier studies it was shown 
that the hopping induced MCT relaxation happens even at lower temperatures where the IMCT relaxation is frozen \cite{bridging,sarika_PNAS}.

The effect of hopping can also be introduced in IMCT as parallel channels in the frequency plane \cite{gotze1,bridging,chong}. The expression of $\phi(q,s)$ can be written as,
\begin{eqnarray}
 \phi(q,s)=\frac{1}{s+K_{hop}+K_{MCT}(q,s)}.
 \label{hop-lap}
\end{eqnarray}
$K_{MCT}(q,s)$ and $K_{hop}$ are the relaxation channels for $\phi(q,s)$. $K_{MCT}(q,s)=\frac{\Omega_q^2}{s+\eta_l(s)}$, where $\eta_l(s)=\gamma +\Omega_q^2\mathcal{L}[\mathcal{M}(q,t)]$. Note that in the expression of $\eta_l$ it is the total memory function with the hopping term which is present whereas in the expression of $\eta_{l}^{id}$ the memory function does not have the hopping term. Thus although $K_{MCT}(q,s)$ and $K_{hop}$ appear as parallel channels, the MCT dynamics is coupled to hopping dynamics. In time plane Eq.\ref{hop-lap} can be written as \cite{bridging},
\begin{eqnarray}
 &&\ddot\phi(q,t)+(\gamma+K_{hop}) \dot\phi(q,t)+(K_{hop}\gamma+\Omega_q^2)\phi(q,t)+\Omega_q^2\nonumber\\
 &\times&\int_0^t{dt'\mathcal{M}(q,t')}
 \times  {[\dot\phi(q,t-t')+K_{hop}\phi(q,t-t')]}=0 .
\label{scheme-2}
\end{eqnarray}
Note that though Eq.\ref{scheme-1} and Eq.\ref{scheme-2} are similar but the memory functions are different for these two schemes.
In an earlier work by one of us we have shown that in spite of this difference qualitatively they describe similar phenomena \cite{bridging}. The advantage of the first
scheme is that in this scheme we can decouple the MCT and the activated contributions. Thus it gives us a handy way to separately calculate the contributions from the diffusive and activated dynamics which is not possible in the other scheme \cite{gotze1,gotze2, chong_pre}.
From now onwards we will refer to the first scheme (from 
Eq.\ref{phitot} and Eq.\ref{scheme-1}) as scheme 1 and the second scheme (Eq.\ref{hop-lap} and Eq.\ref{scheme-2}) as scheme 2.  

We can write similar expressions for the self part of the intermediate scattering function. This is given in Appendix I.

\subsection{Mean squared displacement}

The equation for mean square displacement (MSD) is obtained from the equation of motion for $\phi^s(q,t)$.  The details of the derivation are given in Appendix I. Interestingly we find that in both the schemes although the equations for the intermediate scattering functions  ($\phi^s(q,t)$ and $\phi(q,t)$) are not 
identical, for MSD they are the same and the equation for MSD is given by Eq.\ref{msd_scheme-1},
\begin{eqnarray}
 \frac{\partial}{\partial t}<\Delta r^2(t)>&=&6D_{hop}+6D_0+6D_{hop}D_0\int dt'\mathcal{M}^s(0,t')\nonumber\\
             &-&D_0\int dt'\mathcal{M}^s(0,t')\frac{\partial}{\partial t'}<\Delta r^2(t-t')>.
\label{msd_scheme-1}
\end{eqnarray}
\noindent
Where 
$D_0=\frac{k_BT}{\gamma}$ and $\mathcal{M}^s(0,t)=\frac{1}{(2\pi)^2\rho}\int{d\bf{k}}k^2C^2(k)S(k)\phi^s(k,t)\phi(k,t)$,
where $\phi^s(k,t)=\phi^s_{MCT}(k,t)\phi^s_{hop}(k,t)$ and $\phi^s_{hop}(q,t)=\exp(-K^s_{hop}t)$. $K^s_{hop}$ 
is written as
$K^s_{hop}=\frac{v_0}{v_p}\frac{q^2l^2}{(1+q^2l^2)}P_{hop}(\Delta F)=\frac{q^2l^2}{1+q^2l^2}D_{hop}$, where 
$D_{hop}=\frac{v_0}{v_p}P_{hop}(\Delta F)$ and in $q \rightarrow 0$ limit, $K_{hop}^s=q^2D_{hop}$.

From the above 
equation (Eq.\ref{msd_scheme-1}) we can not a priori separate out the MCT and hopping contributions to MSD. However the MCT contribution to MSD can be evaluated from the equation of motion for $\phi^s_{MCT}(q,t)$. The equation for $<\Delta r^2_{MCT}(t)>$ can be written as,
\begin{eqnarray}
 \frac{\partial}{\partial t}&<&\Delta r_{MCT}^2(t)>=6D_0\nonumber\\
 &-&D_0\int{dt'\mathcal{M}^s(0,t')\frac{\partial}{\partial t'}<\Delta r_{MCT}^2(t-t')>}.
 \label{msd-mct}
\end{eqnarray}
If we now compare Eq.\ref{msd_scheme-1} and Eq.\ref{msd-mct}, we find that when we replace $<\Delta r^2(t)>=<\Delta r_{MCT}^2(t)>+6D_{hop}t$ in Eq.\ref{msd_scheme-1}  we recover Eq.\ref{msd-mct}. This implies that in diffusion the MCT and hopping dynamics are additive. We can now estimate the independent contributions from MCT and hopping dynamics to diffusion ($D_{MCT}=\lim_{t\to\infty} \frac{<\Delta r_{MCT}^2(t)>}{6t}$).

Note that similar to structural relaxation the MCT contribution to the diffusion is coupled to hopping via the memory function.

\subsection{SE breakdown using earlier model}

In this section we investigate the SE relation as predicted by the Unified theory (using schemes 1 and 2, Eq.\ref{scheme-1} and Eq.\ref{scheme-2})
and compare the theoretical predictions with 
experimental data. As discussed in the Introduction, at high temperatures the diffusion coefficient and shear viscosity are related by the Stokes-Einstein relation \cite{hansen_mcdonald,landau_fluid_mech}: D=$\frac{k_BT}{C\eta R}$, where C is a constant and
R is the radius of the moving particle.
Usually in experiment SE validity is measured by calculating the ratio $D\eta T^{-1}/D_0\eta_0T_0^{-1}$, where D is diffusion coefficient,
$\eta$ is the viscosity and T is the temperature and $D_0$, $\eta_0$ are the values at high temperature ($T_0$).  
When this ratio 
grows from unity, SE relation fails. Here for numerical studies we take the ratio $D\tau^{s}/D_0\tau^{s}_0$, i.e. we replace $\eta/T$ by $\tau^{s}$, where
$\tau^{s}$ is the $\alpha$ relaxation time for the self intermediate scattering function. $\tau^{s}$ is defined as the time where $\phi^{s}(q=7.0,t)=0.1$. The diffusion coefficient is calculated from
long time asymptote of the mean squared displacement (MSD).

\begin{figure}[ht!]
\centering
\includegraphics[width=0.47\textwidth]{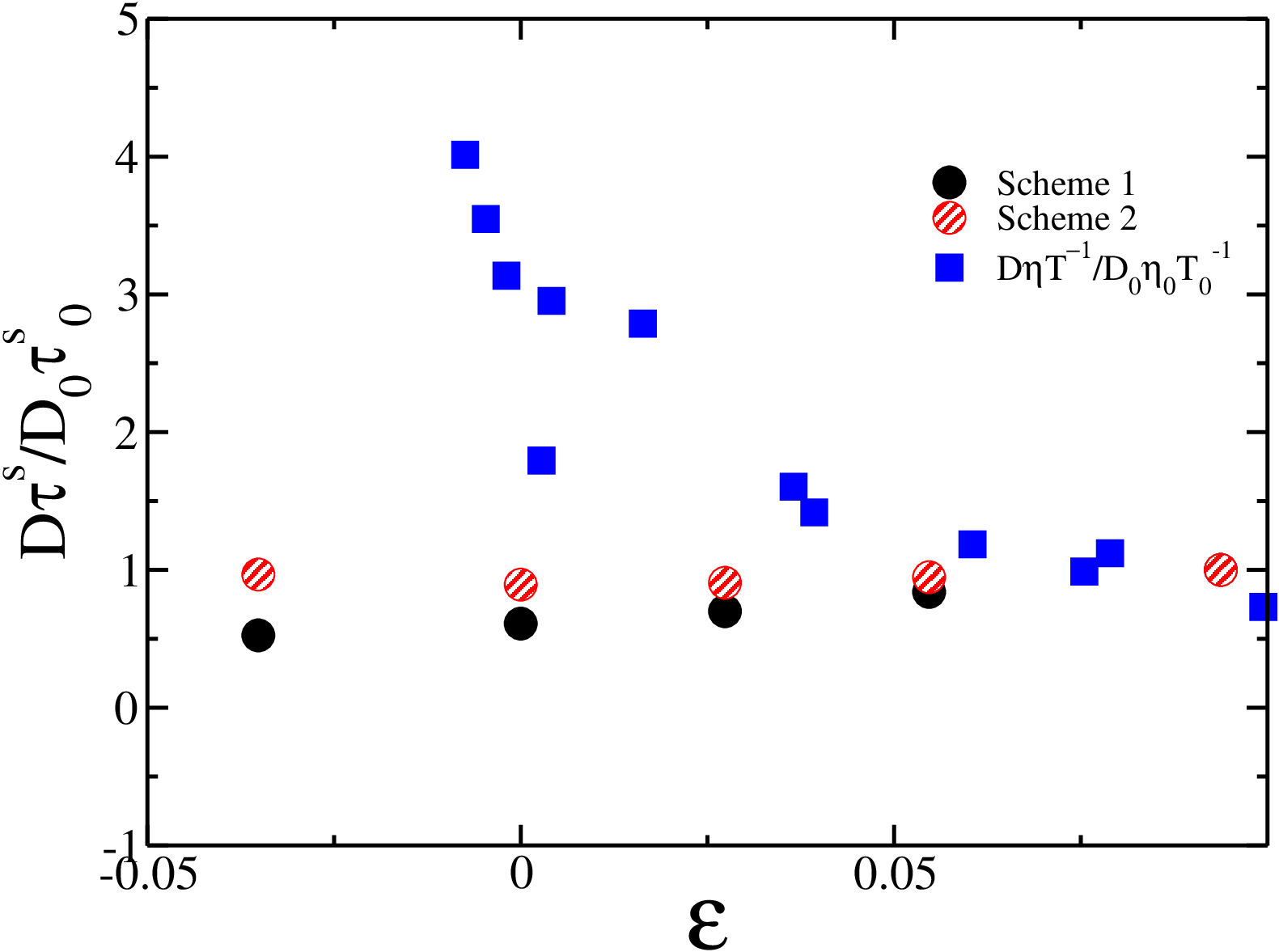}
\caption{ \it{ $D\tau^s/D_0\tau_0^s$ against reduced temperature $\epsilon=(T-T_c)/T_c$ for Salol using scheme 1 and scheme 2. Here the relaxation time is calculated at first peak of S(q). For comparison we have plotted the experimental result \cite{salol-data}. Both the schemes fails to capture SE breakdown.}}
\label{scheme-1and2}
\end{figure}

In Fig.\ref{scheme-1and2} we plot the scaled product of diffusivity, $D$ and relaxation time, $\tau^{s}$ against temperature for Salol system (see Appendix II for the calculational details).  We find that neither of the two schemes are able to capture the Stokes-Einstein breakdown. Note that in an earlier study by Chong using an extended MCT formalism (similar to scheme 2) SE breakdown has been reported \cite{chong_pre}. However the author himself commented that near $T_{g}$ the breakdown is weak. 

In Chong's work the relaxation time is calculated from Eq.4 in Ref.\cite{chong_pre}. This equation is similar to our Eq.\ref{hop-lap}. The MSD in Chong's work is calculated from the self part of Eq.1a ({\it i.e.} Eq.B1) in Ref.\cite{chong_pre} which is similar to Eq.\ref{msd-mct} of this work. Note that Eq.\ref{msd-mct} is not the time plane counterpart of the self part of Eq.\ref{hop-lap}. Similarly
Eq.1a in Ref.\cite{chong_pre} is not the timeplane counterpart of Eq.4 in Ref.\cite{chong_pre}. Thus in Chong's work the equation for MSD just like our Eq.\ref{msd-mct}, only provides the MCT contribution to the diffusion. The equation of motion for MSD which is consistent with Eq.\ref{hop-lap} and its time plane counterpart, Eq.\ref{scheme-2} is Eq.\ref{msd_scheme-1}. Note that in Eq.\ref{msd_scheme-1} the activated motion also contributes to the diffusion. One may argue that the contribution from the activated motion is really small so it does not make any difference if diffusion is calculated from the total MSD or the MCT part of the MSD.  However, at high temperatures the activation barriers are small and the activated process is fast. Thus $D_{hop}$ at high temperatures contribute substantially to the total diffusion and cannot be neglected. At low temperatures the senario is different and the contribution of $D_{hop}$ to the total dynamics is small. This will be discussed in details in the next section. Thus we expect that the degree of decoupling obtained in the earlier work \cite{chong_pre} will decrease when the full diffusion value is taken into consideration. This will further weaken the already weak decoupling. 
Hence it appears that any form of extended MCT including the Unified theory is not successful in describing the SE breakdown.

\subsection{Extended Unified theory}
In this section we make an attempt to extend the Unified theory using the concepts of continuous time random walk.
First we discuss the existing concepts of CTRW. Next we discuss how these concepts can be incorporated in the Unified theory.
Note that in the CTRW description if there is a single waiting time distribution of displacement,
$\psi(t)$ then the dynamic structure factor is given by \cite{montrol_wises},
\begin{eqnarray}
 \phi^{single}_{CTRW}(q,s)=\frac{1-\hat \psi (s)}{s}\frac{1}{1-\hat \psi (s) f(q)}.\nonumber\\
 \label{single_wt}
\end{eqnarray}
Where $f(q)$ is the form factor which is related to the length of displacement.
If the waiting time distribution is exponential, $\psi(t)=\frac{1}{\tau}\exp (-t/\tau)$, where $\tau$ is the timescale of the 
distribution, then Eq.\ref{single_wt} in time plane becomes,
\begin{equation}
 \phi^{single}_{CTRW}(q,t)=\exp (-t(1-f(q))/\tau).
\end{equation}

However there are systems where there are more than one kind of dynamics present.  
Like in supercooled liquids the activated and the MCT like diffusive dynamics. We can then assume that there are two different waiting time distributions $\psi_1^0(t)$ and $\psi_2^0(t)$ related to the two different processes \cite{ctrw_2dis}. 
The bare waiting time distributions get modified due to the presence of each other. The modified waiting time distributions can be 
written as,
\begin{eqnarray}
 \psi_1(t)=\psi^0_1(t)(1-\int_0^t \psi^0_2(t')dt')\nonumber\\
 \psi_2(t)= \psi^0_2(t)(1-\int_0^t \psi^0_1(t')dt').
\end{eqnarray}
According to CTRW formalism \cite{tachiya_bagchi_seki} where there are two types of distributions, we can write the structural relaxation as,
\begin{eqnarray}
 \phi^{double}_{CTRW}(q,s)&=&\frac{1-\hat\psi_1(s)-\hat \psi_2(s)}{s}\nonumber\\
                &\times&\Big(\frac{1}{1-f_1(q)\hat\psi_1(s)-f_2(q)\hat\psi_2(s)}\Big).
\label{two_wt}
\end{eqnarray}
If both the bare waiting time distributions are exponential {\it i.e} $\psi^0_{i}=\frac{1}{\tau_{i}}\exp(-t/\tau_{i})$, where $i=1,2$, then  
in time plane the structural relaxation is given by the product: 
\begin{eqnarray}
 \phi^{double}_{CTRW}(q,t)&\simeq& \exp(-t(1-f_1(q))/\tau_1)\nonumber\\
 &\times& \exp(-t(1-f_2(q))/\tau_2).
\label{ctrw_prod}
\end{eqnarray}
Now if we identify the first as activated and the second as diffusive contribution then the above equation can be written as,
\begin{equation}
 \phi^{double}_{CTRW}(q,t)=\phi_{hop}(q,t)\times \phi_{MCT}(q,t).
 \label{ctrw_prod-2}
\end{equation}

Note that this CTRW formalism gives the same result as given by Eq.\ref{scheme-1} and thus similar to that given by Eq.\ref{scheme-2}. This observation is extremely important in understanding the failure of extended MCT like theories in explaining the SE breakdown.

The origin of SE breakdown lies in the fact that the structural relaxation is dominated by the slow dynamics whereas the diffusion is dominated by the fast dynamics leading to a decoupling between them. As the difference in timescale between the fast and the slow dynamics increases this decoupling becomes stronger. It has been earlier shown that the dynamics described by Eq.\ref{ctrw_prod} is dominated by the fast timescale \cite{tachiya_bagchi_seki}. Hence for any equation which has a similar structure (like  Eq.\ref{scheme-1} and Eq.\ref{scheme-2}), the relaxation dynamics will be predicted by a timescale which is closer to the fast process.
 As the difference between the fast and the slow timescales
 become wider this effect of the dominance of the fast process becomes stronger. Thus for theories where the structural relaxation is given by an expression similar to Eq.\ref{ctrw_prod}, in the intermediate regime although there might be weak decoupling between the diffusion and the structural relaxation, the decoupling will not grow with decrease in temperature. This is what has been observed in an earlier study \cite{chong_pre}.

Since we could show that CTRW theory and Unified theory predicts similar expressions for structural relaxation we can now apply concepts usually used in CTRW theory to the Unified theory.
According to renewal theory we can define two different timescales for a single waiting time distribution \cite{renewal_theory}. They are actually different moments of the distribution. 
One is related to the second moment and is called the persistent time, $<\tau_{p}>$  and the other is the first moment and is called the exchange time, $<\tau_{x}>$. When the waiting time distribution describes a Poisson process {\it i.e.} it is an exponential function then the persistent time is same as the exchange time. However if the waiting time distribution has a tail then persistent time which is the time of the first jump is bigger than the exchange time which describes the subsequent jumps. Thus the first jump takes place at a much longer time than the average value of the 
jumps. This concept has been extensively used in describing the dynamics in kinetically constrained model (KCM) \cite{Berthier3,Berthier9,pinaki21}. According to the renewal theory the 
structural relaxation is given by,
\begin{eqnarray}
\phi_{CTRW}^{renewal}(q,t)=P(t)+\int_0^t dt' p(t')\phi_{diff}(q,t-t').
\label{phi_ctrw}
\end{eqnarray}
Where $P(t)=1-\int_0^t p(t')dt'$ is the persistence function \cite{Berthier3,Berthier9,pinaki21} and $p(t)$ is the persistence time
distribution
and $\phi_{diff}(q,t)$  is the diffusive part of the structural relaxation. With the help of the relation between $P(t)$ and $p(t)$
we can write the above equation as,
\begin{eqnarray}
\phi_{CTRW}^{renewal}(q,t)=P(t)-\int_0^t dt' \dot P(t')\phi_{diff}(q,t-t').
\label{phi_ctrw2}
\end{eqnarray}
Where $\frac{dP(t)}{dt}=-p(t)$.

We now extend this concept and apply it to the CTRW theory with two different waiting time distributions. 
If there are two waiting time distributions, then assuming both are exponential in nature, the total waiting time distribution can be written as,
\begin{eqnarray}
 {\cal P}_{waiting}(t)=\frac{1}{2\tau_1}exp(-t/\tau_1)+\frac{1}{2\tau_2}exp(-t/\tau_2).
 \label{waiting_time_dis}
\end{eqnarray}
The exchange time which is the first moment of the probability distribution function is,
\begin{eqnarray}
 <\tau_x>=\frac{1}{2}(\tau_1+\tau_2).
\end{eqnarray}
The persistence time can be written as \cite{renewal_theory,Chandler},
\begin{eqnarray}
 <\tau_p>=\frac{<\tau_x^2>}{2!<\tau_x>}=\frac{(\tau_1^2+\tau_2^2)}{(\tau_1+\tau_2)}.
\end{eqnarray}
Except when the two timescales ($\tau_{1}$ and $\tau_{2}$ ) are the same, the persistent time is always greater than the exchange time.
If $\tau_2 <<\tau_{1}$  then $<\tau_p> \simeq \tau_1$. 

Thus we show that for such systems as long as the two timescales are different the persistent time is always bigger than the exchange time even when each of them are Poisson process (given by exponential functions). When the two waiting time distributions are widely apart then $\tau_{p}$ is closer to the slower process. 
Next we map the functions $P(t)$ and 
$\phi_{diff}(q,t)$ with the functions in the Unified theory. At low temperatures the activated dynamics is the slowest dynamics
in the system. Hence the persistence function can be related to the 
activated dynamics, 
\begin{equation}
 P(t)=\exp(-P_{hop}(\Delta F)t).
\end{equation}
Where $P_{hop}(\Delta F)=\frac{1}{\tau_0}\exp(-\Delta F/k_BT)$.

Next we can identify the diffusive part of the structural 
relaxation  $\phi_{diff}(q,t-t')$ with $\phi_{CTRW}^{double}(q,t)$ and thus with Eq.\ref{phitot}.  Eq.\ref{phi_ctrw2} when recast in terms of the Unified theory can be written as,
\begin{eqnarray}
 \phi^{renewal}_{CTRW}(q,t)&=& P(t)-\int_0^t \dot P(t')\phi^{s}(q,t-t')dt'.
 \label{ext_mct_ctrw}
\end{eqnarray}
Where $\phi_{diff}(q,t)=\phi^{s}(q,t)=\phi^{s}_{hop}(q,t)\phi^{s}_{MCT}(q,t)$ (Appendix I). 
Here we assume that the time scale for the first jump comes from the waiting time distribution of activation and subsequent jumps from either MCT or activated dynamics. Similar to that discussed in an earlier work \cite{pinaki21} the diffusive part of the dynamics is not effected by the renewal theory and is same as given in the first part of this article

\begin{figure}[ht!]
\centering
\includegraphics[width=0.47\textwidth]{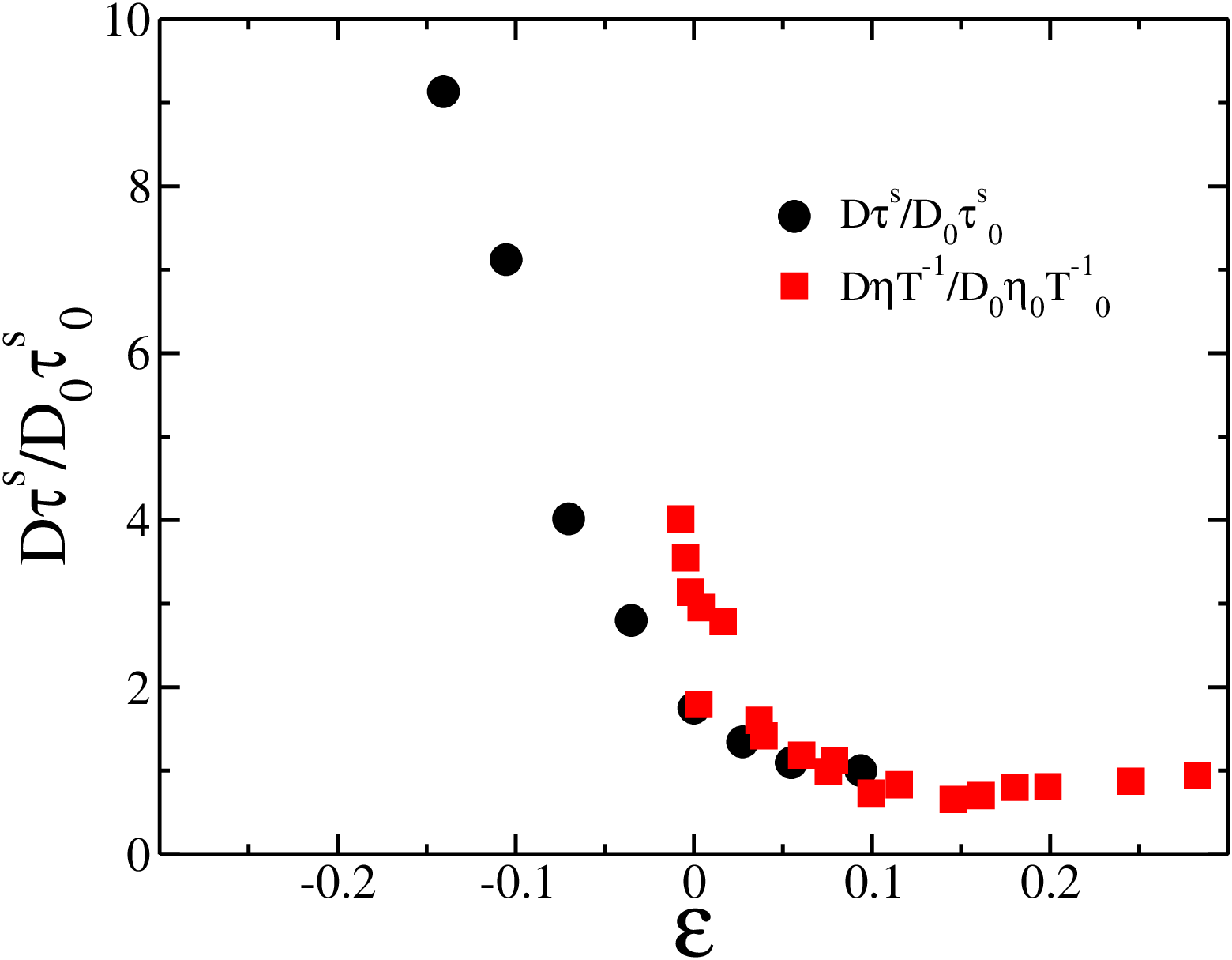}
\caption{ \it{ SE breakdown plot for Salol system. Similar to Fig.1, $D\tau^s/D_0\tau_0^s$ is plotted against reduced temperature $\epsilon=(T-T_c)/T_c$. Here the relaxation time is calculated using the extended Unified theory (Eq.\ref{ext_mct_ctrw}). $D\eta T^{-1}/D_0\eta_0T_0^{-1}$ are experimental data taken from Ref. \cite{salol-data}. The extended Unified theory  is able to capture SE breakdown.}}
\label{se-break}
\end{figure}

\section{Results}
\subsection{Break down of SE relation}
We now obtain the $\tau^{s}$ from the newly derived expression (Eq.\ref{ext_mct_ctrw}). 
Thus $\tau^{s}$ is defined as the time where $\phi^{renewal}_{CTRW}(q=7.0,t)=0.1$. 
In Fig.\ref{se-break} we plot
 the ratios, $D\tau^{s}/D_0\tau^{s}_0$ ($D_{0}$ and $\tau^{s}_0$ are the high temperature diffusion value and relaxation time respectively, calculated at the onset temperature T=280K). For comparison, the experimental data are shown as
 $D\eta T^{-1} /D_0 \eta_0 T_{0}^{-1}$ \cite{salol-data}. Note that the MSD formalism remains unchanged as $\phi_{diff}(q,t)=\phi^{s}_{hop}(q,t)\phi^{s}_{MCT}(q,t)$ contributes to the diffusive dynamics.
 In this figure we find that our numerical result is consistent with the 
experimental data and is able to show the SE break down. 
Our study clearly shows that the present formalism predicts a strong decoupling between the diffusion and structural relaxation. The diffusion coefficient is about 10 times faster than that predicted by SE relation. Also note that the nature of the plot does not show a saturation of the decoupling at lower temperatures. In our study the decoupling value never reaches 100 as observed in experiments on OTP \cite{ediger_jpcb}. This may be because for OTP the barrier height for the activation dynamics and other parameters are different from that of Salol. According to RFOT theory the rate at which the barrier height increases with temperature varies with system \cite{lubchenko2003barrier} and in our theory this effects the temperature dependence of the relaxation time and thus the degree of decoupling.


\begin{figure*}[ht!]
\centering
\subfigure{
\includegraphics[width=0.45\textwidth]{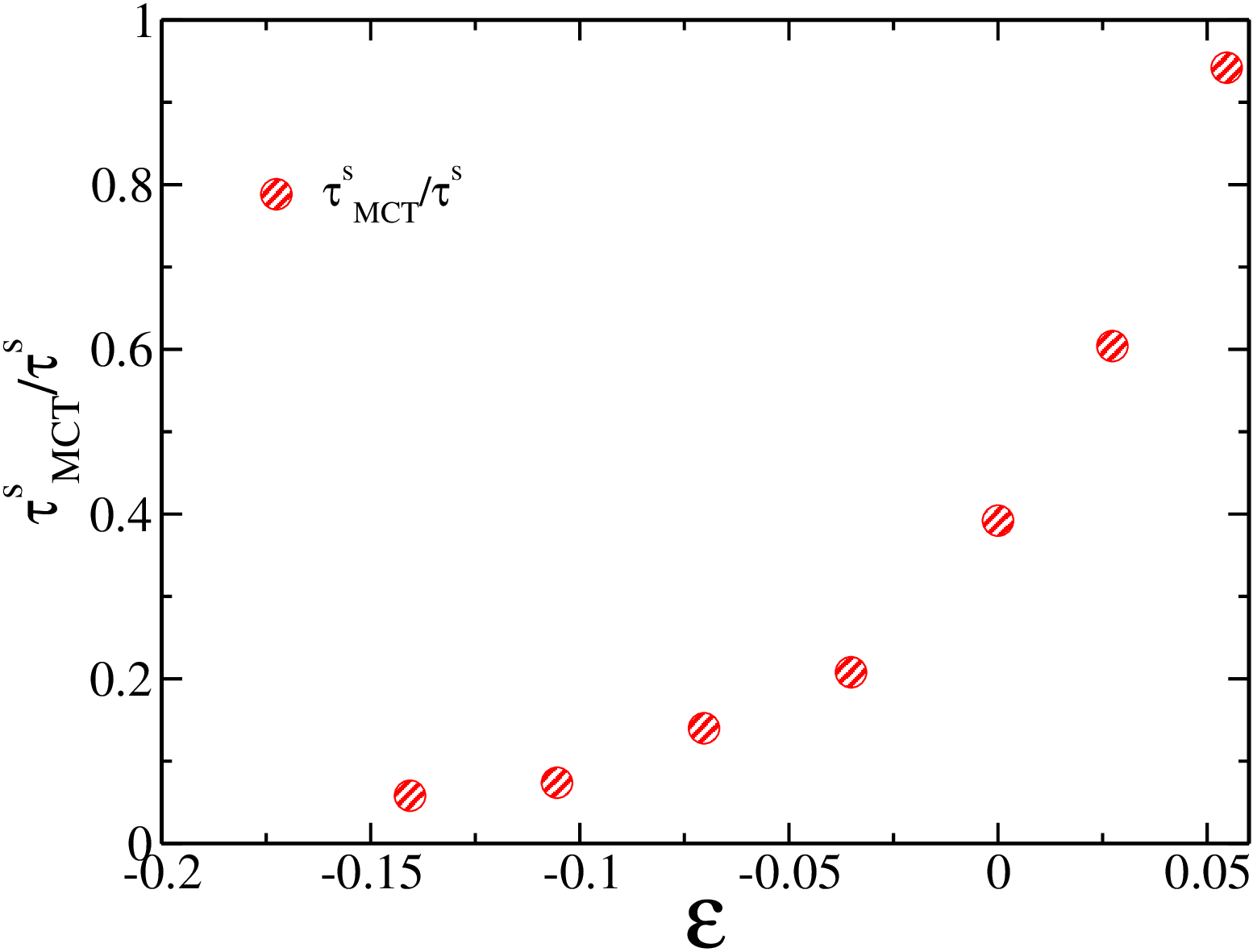}}
\subfigure{
\includegraphics[width=0.45\textwidth]{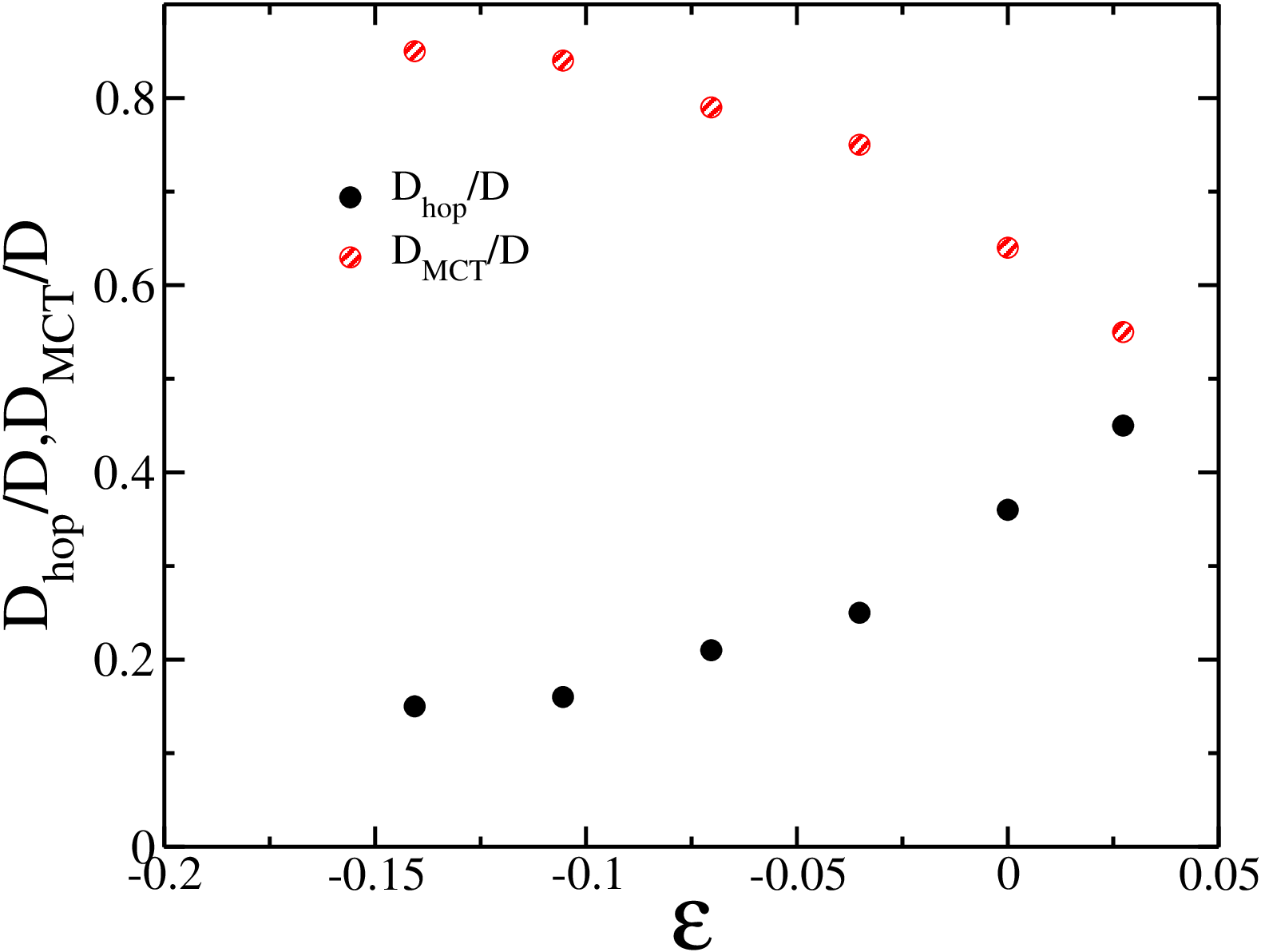}}
\caption{ \it{(a) Comparison of the MCT dynamics to the total relaxation time. $\frac{\tau^{s}_{MCT}}{\tau^{s}}$ is plotted against $\epsilon=(T-T_{c})/T_{c}$. At high
temperature the dynamics is similar to MCT, whereas at low temperature the total dynamics is much slower than the MCT dynamics and is dominated by the activation process. (b) Estimation of the independent contributions from activated and MCT like dynamics to the diffusion.  $D_{hop}/D$ and
$D_{MCT}/D$ are plotted against $\epsilon=(T-T_{c})/T_{c}$. At high
temperature the contribution of hopping and MCT are equal and at low temperature diffusion is dominated by MCT dynamics.}}
\label{tau_contribution}
\end{figure*}

In order to understand the origin of SE breakdown we analyze the different contributions from MCT and activated dynamics to relaxation
time and diffusion. Note that by construction, our total relaxation time will be close to the activation (hopping) timescale. In Fig.\ref{tau_contribution}(a) we plot $\frac{\tau^{s}_{MCT}}{\tau^{s}}$. Where $\tau^{s}_{MCT}$ is 
obtained by solving self part of Eq.\ref{phi-prod} (Eq.\ref{imct_fskt} in Appendix I). 
The present formalism predicts that the dynamics at high temperatures is similar to the MCT dynamics and at low temperatures the MCT dynamics is much faster than the structural relaxation. We do a similar analysis for the diffusion coefficient. In Fig.\ref{tau_contribution}(b) we plot $D_{hop}/D$ and
$D_{MCT}/D$. These ratios provide us the contribution of the activated and MCT dynamics to the diffusion.
We find that at high temperatures when the barriers are small and thus the activation process is more frequent, both
activation and MCT contributes equally to the diffusion ($\frac{D_{hop}}{D}\simeq \frac{D_{MCT}}{D}\simeq 0.5$). At this temperature
may be it is difficult to even differentiate between the two processes. However as temperature is lowered the activation becomes a rare event
and the diffusion is determined by the MCT like dynamics 
($\frac{D_{hop}}{D}\to 0, \frac{D_{MCT}}{D}\to 1$). Thus our present formalism clearly shows that at low temperatures the dynamics
is determined by the activated process and diffusion by MCT process and this is responsible for the SE breakdown shown in Fig.\ref{se-break}. This decoupling is similar to that explained in KCM where it was shown that the persistent and the exchange waiting time distributions are similar at high temperatures and they decouple as the temperature is lowered \cite{Chandler}. According to KCM this decoupling is related to the SE breakdown.

\begin{figure*}[ht!]
\centering
\subfigure{
\includegraphics[width=0.45\textwidth]{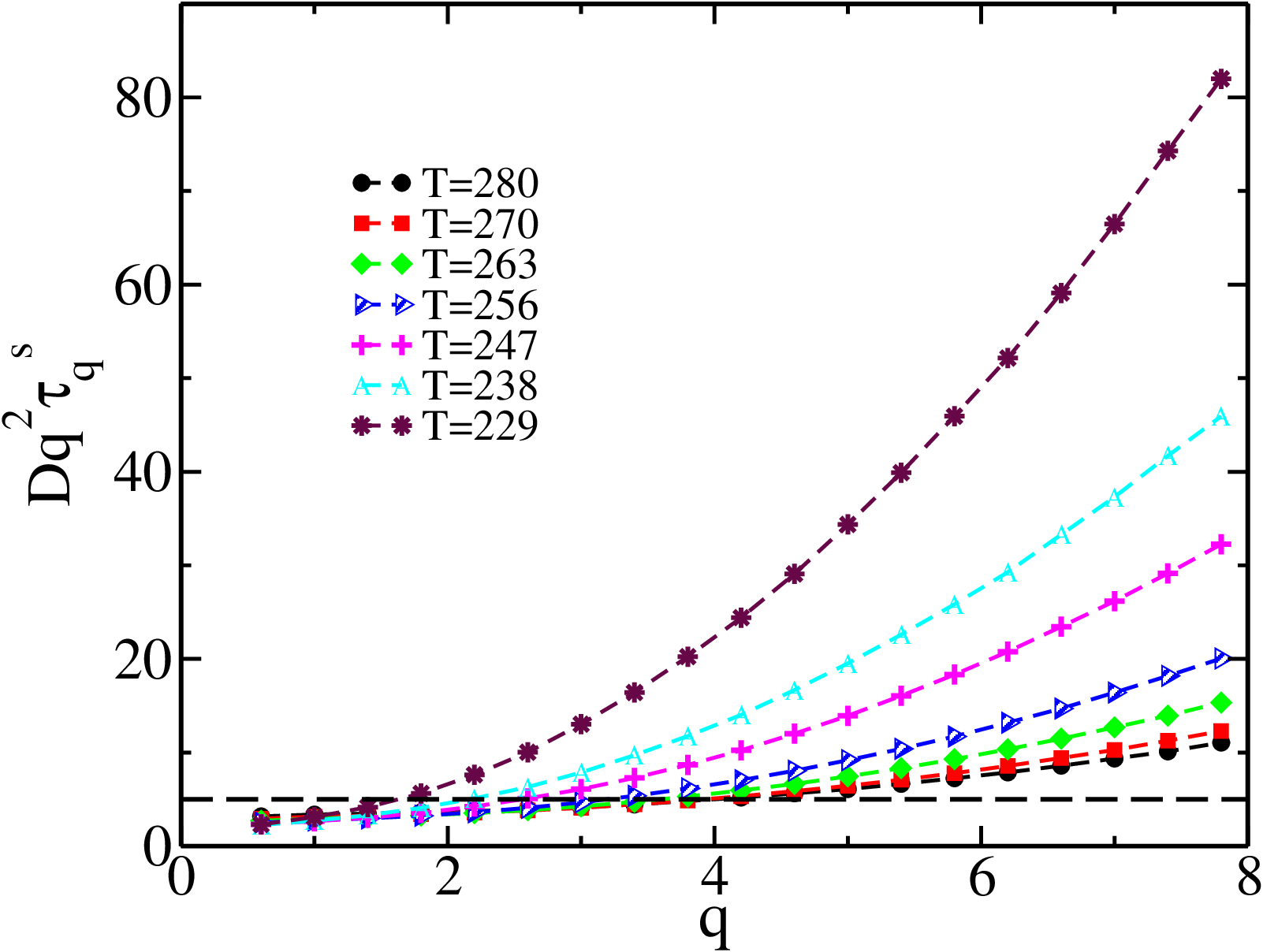}}
\subfigure{
\includegraphics[width=0.45\textwidth]{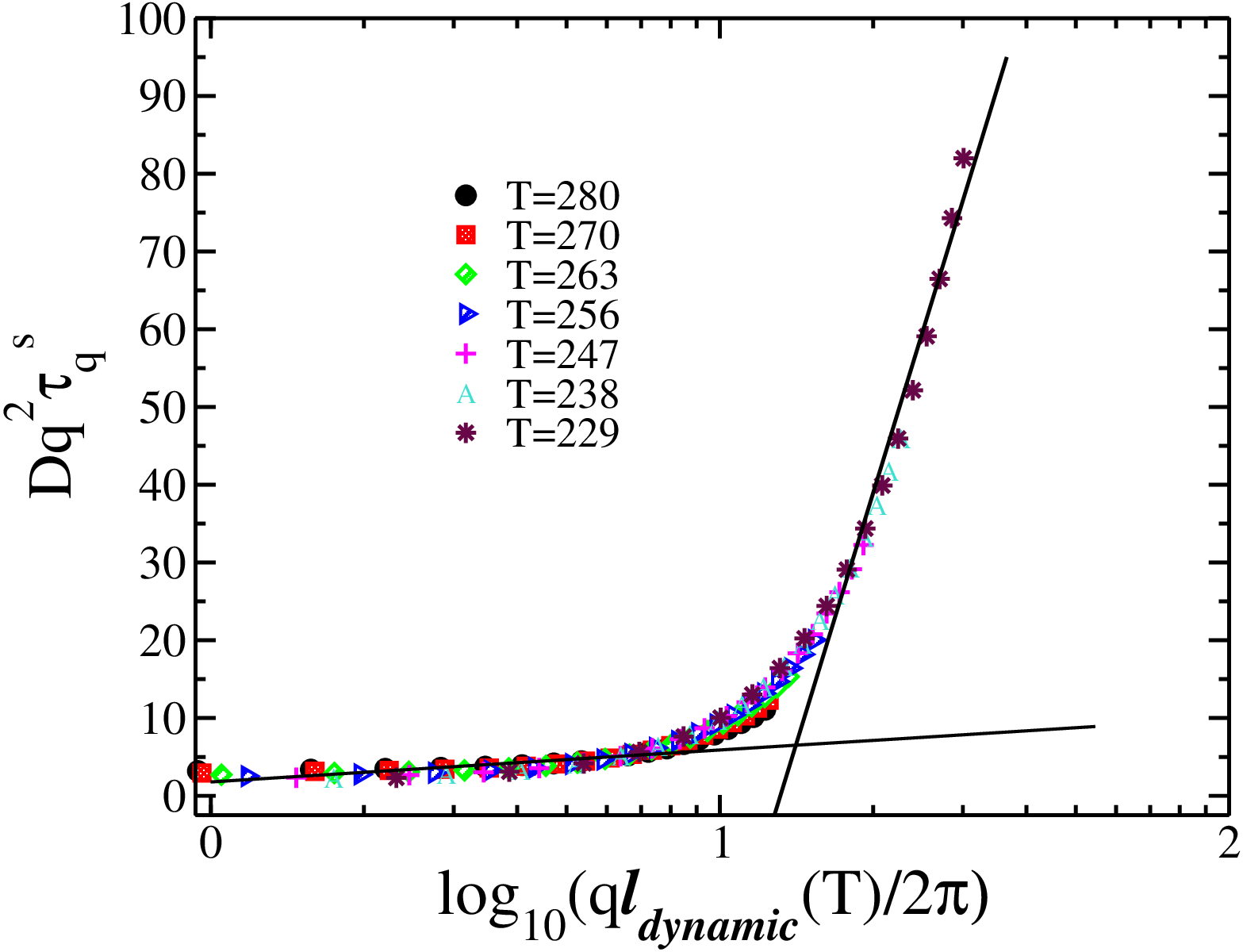}}
\caption{ \it{(a)$Dq^2\tau_q^s$ plotted against wavenumber q for different temperatures. At each temperature the plot shows a Fickian to non-Fickian transition. Since the transition is continuous, to consistently describe a temperature dependent transition wavenumber $q^{*}(T)$, we draw a horizontal dashed line. $q^{*}(T)$ is the wavenumber where $Dq^2\tau_q^s$ cuts the dashed line. (b)D$q^2\tau^s_q$ follows a master plot when plotted against $ q l_{dynamic}(T)/2 \pi $, where the dynamic lengthscale $l_{dynamic}=2 \pi/q^{*}$. }}
\label{dtau_q}
\end{figure*}

\subsection{Growing length scale }
\begin{figure}[ht!]
\centering
\includegraphics[width=0.47\textwidth]{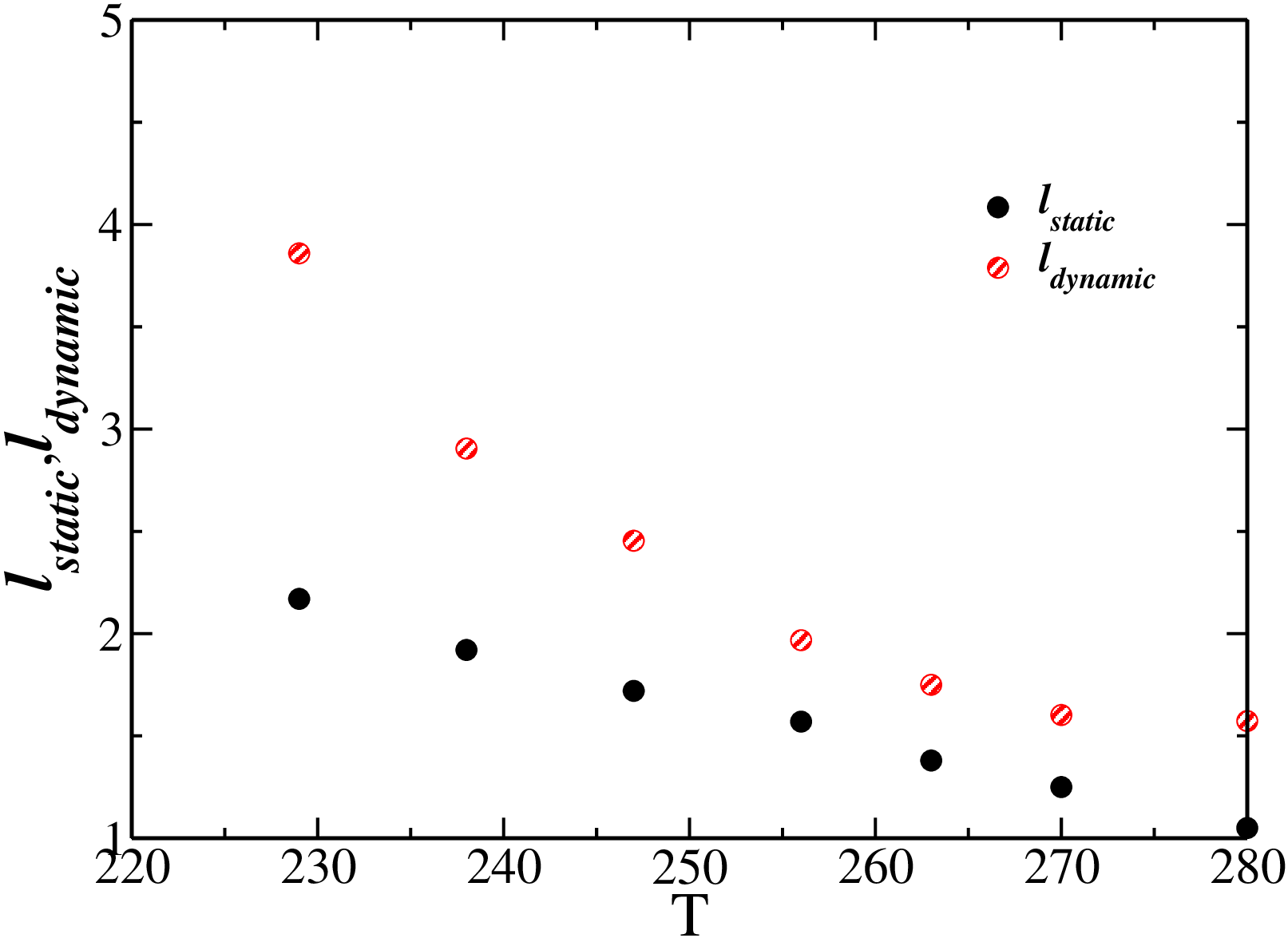}
\caption{ \it{The plot of static and dynamic lengthscales $l_{static}$ and $l_{dynamic}$ against temperature. The dynamic correlation length
grows more than the static correlation length.}}
\label{static_dynamic}
\end{figure}

The relaxation time obtained from Eq.\ref{ext_mct_ctrw} is dependent on wave number, $q$. At small $q$ it is expected to show a Fickian behaviour
{\it i.e} $Dq^2\tau_q^{s}\sim$ constant. In Fig.\ref{dtau_q}(a) we plot the $Dq^2\tau_q^s$ vs $q$. The plot shows that $Dq^2\tau_q^s$ is a constant only at small $q$ and shows a strong $q$ dependence at higher wavenumbers.
This Fickian to non-Fickian transition is temperature dependent. At low temperatures this transition happens at smaller $q$ values which means that the length scale over which the system becomes diffusive, increases with decrease in temperature. 
It is also believed that as the cooperatively rearranging region in the system grows the Fickian behaviour 
sets in at longer a lengthscale. Thus the transition of $\tau_q^{s}$ from Fickian to non-Fickian
does provide us a measure of the length scale of the CRR. 
This transition is gradual and in order to obtain a temperature dependent length scale, we choose the $q^*$ value where $Dq^2\tau_q^s=5$.
The corresponding length scale is $l_{dynamic}=\frac{2\pi}{q^*}$. Next we show that if we scale
the x axis by $l_{dynamic}$ then the plots in Fig.\ref{dtau_q}(a) shows a data collapse. $Dq^2\tau_q^s$ shows a clear 
transition from Fickian to non-Fickian regime. We call this $l_{dynamic}$ the dynamic correlation length and in 
Fig.\ref{static_dynamic} we plot it with respect to temperature. We also plot the static correlation length $l_{static}$ which
is obtained from the RFOT. $l_{static}$ is the length where $\frac{\partial \Delta F}{\partial r}\simeq 0$ and note that this is 
an input to the theory. Thus our formalism shows that the dynamic correlation length grows more than the static correlation length
which has been observed in experimental studies \cite{berthierEPL24}. In our study we do not have a change in shape of 
the CRR \cite{kob_dynamiclengthscale,rajeshCRR} and thus our $l_{dynamic}$ is monotonic with temperature.

\section{Summary}

 At high temperatures the dynamics of a liquid is known to follow the Stokes-Einstein relation. According to the SE relation the diffusion and the structural relaxation timescale/viscosity of the liquid are coupled and follow each other. However the dynamics of the system changes in the supercooled liquid regime, where along with diffusive dynamics there are activated motions. One of the consequences of this change in dynamics is the breakdown of the SE relation.  
The Unified theory (MCT+RFOT) which was developed earlier was successful in extending the MCT like diffusive dynamics till a much lower temperature where the MCT failed \cite{bridging,sarika_PNAS}. It predicted that there is a hopping induced diffusion. With the success of the theory in extending the MCT to low temperatures it becomes imperative to check if the other properties of the dynamics like the SE breakdown can be predicted from this theory. 

In order to study the SE breakdown we need the knowledge of both structural relaxation timescale and the diffusion coefficient. The structural relaxation had already been developed earlier \cite{bridging,sarika_PNAS}. Here we derive the equation for the MSD and from the long time limit of MSD we obtained the diffusion coefficient. The product of the diffusion coefficient and the relaxation time when plotted against temperature does not show any decoupling between them thus showing that the Unified theory is incapable of predicting the SE breakdown. We next investigate the origin of this failure.
 
We show that the formulation of the Unified theory can also be obtained from continuous time random walk formalism. If in CTRW we consider that there are two kinds of dynamics (like diffusive and activated), each having their own waiting time distributions then under the approximation of exponential waiting time distribution the CTRW predicts that the total structural relaxation dynamics is given by the product of the independent structural relaxations of the two different dynamics \cite{tachiya_bagchi_seki}. However, it has been shown that in such case the total dynamics is dominated by the fast motion \cite{tachiya_bagchi_seki}. This is precisely what we obtain from the Unified theory. It predicts that both the structural relaxation and the diffusion are determined by the MCT like dynamics and thus they never decouple. This leads to the failure of the Unified theory in predicting the SE breakdown. Once we know the origin of the failure our next exercise is to extend the theory such that it can predict the SE breakdown.

Note that within the CTRW framework according to the renewal theory, if the waiting time distribution has a long tail then the persistent time is larger than the exchange time. Hence the first jump takes place at a much longer time compared to the average jump rate \cite{renewal_theory,Chandler}. We show here that for a system having two processes and thus two waiting time distributions even if each of them are exponential the 
persistent time is larger than the exchange time. When the timescales of these two processes are widely different then the persistent time is same as the slower process. Since we show that the Unified theory can also be explained from the CTRW formalism we next apply this renewal theory to the existing Unified theory. In the extended Unified theory we assume that the first jump takes place via the slower activated dynamics and the subsequent jumps take place either via the MCT like dynamics or the activated dynamics. 
Following earlier studies we assume that the diffusive dynamics is not effected by this renewal process and is given by the product of the two independent dynamics \cite{berthierEPL23}. We next show that within the extended Unified theory we can predict the SE breakdown and the degree of decoupling is closer to the experimental observations and much stronger than earlier theoretical prediction \cite{chong_pre}.
We further do an analysis of the independent contributions from MCT and activated dynamics to the relaxation timescale and the diffusion. We show that at high temperatures (just below onset) the dynamics is similar to the MCT dynamics and as the temperature is lowered there is a crossover and the dynamics is determined by activated motions. The diffusion on the other hand shows a different scenario. At high temperatures we  
find that both MCT and activated dynamics contributes equally to the diffusion value. This is because at high T the activation barriers are small and the activated dynamics is fast. Thus it contributes to the diffusion process. However at low temperatures the activated dynamics becomes slow and the diffusion is determined by the MCT like dynamics. This leads to the decoupling of the structural relaxation and the diffusion and thus to SE breakdown.

Finally we decided to extract a dynamical lengthscale from our extended Unified theory. We know that only at large lengthscale the dynamics is Fickian (diffusive). This Fickian to non Fickian transition, usually measured by the wavenumber dependence of the relaxation time, provides us a lengthscale beyond which dynamics is diffusive. This can also be considered as the lengthscale beyond which the dynamics is randomized and thus connected to the length of a cooperatively rearranging region. We show that as the temperature is decreased the lengthscale increases. The wavenumber dependence of the relaxation times show a master plot when the wavenumber is scaled by this dynamic lengthscale, predicting a temperature independent Fickian to non-Fickian transition. The dynamic lengthscale is found to grow faster than the static lengthscale. However since in our study we do not incorporate a change in the structure of the CRR \cite{rajeshCRR,kob_dynamiclengthscale} the dynamic lengthscale does not show a non-monotonic behaviour.   

Thus our study predicts that all theoretical formulations which work on the extension of the MCT needs to incorporate the renewal theory to predict the correct dynamics of the liquid in the deeply supercooled regime.

\section{Appendix-I}
\subsection{Mean squared displacement}

Following the derivation of Eq.\ref{scheme-1} we can write the equation for the self part of the 
intermediate scattering function,
$\phi^s(q,t)$  in the over damped limit as,
\begin{eqnarray}
 \dot \phi^s(q,t) &+&\frac{(K^s_{hop})^2+\gamma K^s_{hop}+(\Omega^s)^2}{\gamma+2K^s_{hop}}\phi^s(q,t)\nonumber\\
 &+&\frac{(\Omega^s)^2}{q^2(\gamma+2K^s_{hop})}\int dt' \phi^s_{hop}(q,t')\mathcal{M}^s(q,t')\nonumber\\
 &\times&[\dot \phi^s(q,t-t')+K_{hop}\phi^s(q,t-t')]=0.
 \label{phis_scheme-1-a}
\end{eqnarray}

Where $(\Omega^s)^2=q^2k_BT/m$, 
$\mathcal{M}^s(q,t)=\frac{1}{(2\pi)^2\rho}\int {d\bf{k}}({\bf{q.k}})^2 C^2(k)S(k)\phi^s(p,t)\phi(k,t)$,
where $\phi^s(k,t)=\phi^s_{MCT}(k,t)\phi^s_{hop}(k,t)$.
$K^s_{hop}=\frac{v_0}{v_p}\frac{q^2l^2}{(1+q^2l^2)}P_{hop}(\Delta F)=\frac{q^2l^2}{1+q^2l^2}D_{hop}$, where 
$D_{hop}=\frac{v_0}{v_p} P_{hop}(\Delta F)$
and $\phi^s_{hop}(q,t)=\exp(-K^s_{hop}t)$.
Next we derive the equation for mean squared displacement (MSD) from Eq.\ref{phis_scheme-1-a}. In order to do that we need to go to the small
q limit of Eq.\ref{phis_scheme-1-a}.

In the small q we can expand $\phi^s(q,t)\simeq 1-\frac{q^2}{6}<\Delta r^2(t)>$ and 
$\phi^s_{hop}(q,t)\simeq 1-{q^2}D_{hop}t$ ($q\rightarrow0$, $K^s_{hop}=q^2D_{hop}$) and write the above equation as,
\begin{eqnarray}
 &-\frac{q^2}{6}\frac{\partial}{\partial t}<\Delta r^2(t)>
 +\Big(\frac{q^4D_{hop}^2+\gamma q^2D_{hop}+q^2k_BT}
 {\gamma+2q^2D_{hop}}\Big)\nonumber\\
 &\times(1-\frac{q^2}{6}<\Delta r^2(t)>)
+\frac{q^2k_BT}{q^2(\gamma+2q^2D_{hop})}\nonumber\\
&\times\int dt'(1-q^2D_{hop}t')\mathcal{M}^s(q,t')(-\frac{q^2}{6}\frac{\partial}{\partial t'}<\Delta r^2(t-t')>)\nonumber\\
&+\frac{q^2k_BT}{q^2(\gamma+2q^2D_{hop})}\int dt'(1-q^2D_{hop}t')\mathcal{M}^s(q,t')\nonumber\\
&\times q^2D_{hop}(1-\frac{q^2}{6}<\Delta r^2(t-t')>)=0. 
\end{eqnarray}
Now, if we divide by  $q^2$, the above equation becomes,
\begin{eqnarray}
 &-\frac{1}{6}\frac{\partial}{\partial t}<\Delta r^2(t)> +\Big(\frac{q^2D_{hop}^2+\gamma D_{hop}+k_BT}{\gamma+2q^2D_{hop}}\Big)\nonumber\\
 &\times (1-\frac{q^2}{6}<\Delta r^2(t)>)+\frac{k_BT}{\gamma+2q^2D_{hop}}\nonumber\\
 &\times \int dt'(1-q^2D_{hop}t')\mathcal{M}^s(q,t')(-\frac{1}{6}\frac{\partial}{\partial t'}<\Delta r^2(t-t')>)\nonumber\\
&+\frac{D_{hop}k_BT}{\gamma+2q^2D_{hop}}\int dt'(1-q^2D_{hop}t')\mathcal{M}^s(q,t')\nonumber\\
&\times (1-\frac{q^2}{6}<\Delta r^2(t-t')>)=0. 
\end{eqnarray}
In $q\rightarrow 0$ limit the memory function for MSD becomes,
\begin{equation}
 \mathcal{M}^s(0,t)=\frac{1}{(2\pi)^2\rho}\int {d\bf{k}}k^2C^2(k)S(k) \phi^s(k,t)\phi(k,t).
\end{equation}

In $q\rightarrow 0$ limit we get,
\begin{eqnarray}
 -\frac{\partial}{\partial t}&<&\Delta r^2(t)> +6(D_{hop}+\frac{k_BT}{\gamma})\nonumber\\
 &-&\frac{k_BT}{\gamma}\int dt'\mathcal{M}^s(0,t')\frac{\partial}{\partial t'}<\Delta r^2(t-t')>\nonumber\\
 &+&6D_{hop}\frac{k_BT}{\gamma}\int dt'\mathcal{M}^s(0,t')=0.
\end{eqnarray}
From the above equation we can write the mean squared displacement for scheme 1 as,
\begin{eqnarray}
 \frac{\partial}{\partial t}<\Delta r^2(t)>&=&6D_{hop}+6D_0+6D_{hop}D_0\int dt'\mathcal{M}^s(0,t')\nonumber\\
             &-&D_0\int dt'\mathcal{M}^s(0,t')\frac{\partial}{\partial t'}<\Delta r^2(t-t')>.
\label{msd_scheme-1-a}
\end{eqnarray}
Where $D_0=\frac{k_BT}{\gamma}$.


Similarly in scheme 2, we can write the equation for $\phi^s(q,t)$ in overdamped limit as,
\begin{eqnarray}
 \dot\phi^s&(&q,t)+\frac{\gamma K^s_{hop}+q^2k_BT}{\gamma +K^s_{hop}}\phi^s(q,t)\nonumber\\
 &+&\frac{k_BT}{\gamma +K^s_{hop}}\int_{0}^{t}{dt'\mathcal{M}^s(q,t')\dot\phi^s(q,t-t')}\nonumber\\
                    &+&\frac{k_BTK^s_{hop}}{\gamma +K^s_{hop}}\int_{0}^{t}{dt'\mathcal{M}^s(q,t')\phi^s(q,t-t')}=0.
\end{eqnarray}
For small q we can write,
\begin{eqnarray}
 -\frac{q^2}{6}\frac{\partial}{\partial t}<\Delta r^2(t)>+q^2\frac{\gamma D_{hop}+k_BT}{\gamma+q^2D_{hop}}(1-\frac{q^2}{6}<\Delta r^2(t)>)\nonumber\\
+\frac{k_BT}{\gamma+q^2D_{hop}}\int{dt'\mathcal{M}^s(q,t')\frac{\partial}{\partial t'}(1-\frac{q^2}{6}<\Delta r^2(t-t')>)}\nonumber\\
+\frac{q^2k_BTD_{hop}}{\gamma+q^2D_{hop}}\int{dt'\mathcal{M}^s(q,t')(1-\frac{q^2}{6}<\Delta r^2(t-t')>)}=0.\nonumber\\
\end{eqnarray}
In $q\rightarrow 0$ limit, we can write
\begin{eqnarray}
\frac{\partial}{\partial t}&<&\Delta r^2(t)>=6D_{hop}+6\frac{k_BT}{\gamma}\nonumber\\
&+&6\frac{k_BT}{\gamma}D_{hop}\int{dt'\mathcal{M}^s(0,t')}\nonumber\\
 &-&\frac{k_BT}{\gamma}\int{dt'\mathcal{M}^s(0,t')\frac{\partial}{\partial t'}<\Delta r^2(t-t')>}.
\end{eqnarray}

\begin{eqnarray}
\frac{\partial}{\partial t}&<&\Delta r^2(t)>=6D_{hop}+6D_0+6D_{hop}D_0\int{dt'\mathcal{M}^s(0,t')}\nonumber\\ 
 &-&D_0\int{dt'\mathcal{M}^s(0,t')\frac{\partial}{\partial t'}<\Delta r^2(t-t')>}.
\label{msd-hop-a}
\end{eqnarray}
Similarly for IMCT, we can write the equation of motion for $\phi^s_{MCT}(q,t)$ in overdamped limit as,
\begin{eqnarray}
 &&\dot\phi^s_{MCT}(q,t)+\frac{\gamma q^2k_BT}{\gamma }\phi^s_{MCT}(q,t)\nonumber\\
 &+&\frac{k_BT}{\gamma }\int_{0}^{t}{dt'\mathcal{M}^s(q,t')\dot\phi^s_{MCT}(q,t-t')}=0.
 \label{imct_fskt}
\end{eqnarray}
Expanding $\phi^s_{MCT}(q,t)\simeq 1-\frac{q^2}{6}<\Delta r^2_{MCT}(t)>$, we obtain the equation for MSD, 
\begin{eqnarray}
\frac{\partial}{\partial t}&<&\Delta r^2_{MCT}(t)>=6D_0\nonumber\\
&-&D_0\int{dt'\mathcal{M}^s(0,t')\frac{\partial}{\partial t'}<\Delta r^2_{MCT}(t-t')>}.
\label{msd-imct-a}
\end{eqnarray}
\section{Appendix - II}
\subsection{Static structure factor}
In this work, numerical results for Salol are presented. The static structure factor, S(q) is 
calculated within Percus-Yevick (PY) approximation \cite{hansen_mcdonald,landau_fluid_mech}. The range of temperature studied here is T=280 to T=220.
To calculate $S(q)$ for a value of temperature we need to map the hard sphere system to Salol. First of all, we map a Lennard Jones
(LJ) system at $T^{*}_c=1.305$ and $\rho^{*}_c{_{LJ}}=0.99$ to Salol at T=256K, as the IMCT dynamics for the LJ system predicts a transition at that state
point. Then for each temperature between T=280k to 220K we calculate an equivalent $T^*=\frac{1.305T}{256}$.
Keeping $\rho^*_{LJ}=0.99$ for each $T^*$ we calculate equivalent hard sphere density $\rho^*_{HS}$. Using this $\rho^*_{HS}$ in 
Percus-Yevick approximation we calculate the structure
factor for each temperature.

\subsection{Barrier height}
We calculate the barrier height by using RFOT theory. According to RFOT theory, the free energy cost due to formation of the entropy
droplet can be written as \cite{lubchenko2003barrier}, 
\begin{equation}
 F(r)=\frac{\Gamma_K(r) \Gamma_A(r)}{\Gamma_K(r)+\Gamma_A(r)}-\frac{4\pi}{3}r^3TS_c.
\end{equation}
Where $\Gamma_A(r)=4\pi \sigma_Ar^2=4\pi\Delta f a^3 (r/a)^2$ is the surface energy at temperature $T_A$ (where hopping barrier 
disappears), `a' is the length which corresponds to liquid's volume per molecule $a^3=V/N$. V volume and N is the number of molecules. 
$\Delta f$ can be obtained from the relation given below,
\begin{equation}
 \frac{\Delta f}{T_AS_c^A}=4(t/t_K)^{3/2}\frac{1}{\sqrt{1+3(t/t_K)}}.
\end{equation}
Where $t\equiv \frac{T_A-T}{T_A}$ and $t_K\equiv \frac{T_A-T_K}{T_A}$.
 $\Gamma_K(r)=4\pi \sigma_K(r)r^2=4\pi\sigma_0a^2(r/a)^{3/2}$ is the surface energy term at  
temperature $T_K$ (the Kauzmann temperature) \cite{bridging,lubchenko2003barrier} and 
$\sigma_0=\frac{3}{4}(k_BT/a^2)ln((a/d_L)^2/\pi e)$ and $d_L\sim 0.1a$ is the Lindemann length \cite{lindman}.
Configurational entropy is calculated via an empirical 
formula \cite{pnas26} $S_c=S_{fit}(1-T_K/T)$, where $S_{fit}$ is the system dependent fitting parameter. To get the mean barrier height as discussed above, for Salol we fixed $T_A=330 K$ and $T_K=175K$ and $S_{fit}=2.65$ 
\cite{sarika_PNAS,lubchenko2003barrier}. Although we fixed $T_A=330K$, it is known that for Salol the onset temperature is $T=280K$ \cite{sarika_PNAS,beta_experiment}. However this choice of $T_{A}$ can be justified as we find that the radius of the CRR, $l_{static}$ is initially small and only around $T=280K$ is reaches about one particle diameter. 


\section{Appendix - III}

\subsection{CTRW and extended MCT}
Note that in the continuous time random walk description if there is a single waiting time distribution of displacement,
$\psi(t)$ then the dynamic structure factor is given by,
\begin{eqnarray}
 \phi^{single}_{CTRW}(q,s)=\frac{1-\hat \psi (s)}{s}\frac{1}{1-\hat \psi (s) f(q)}.\nonumber\\
 \label{single_wt-c}
\end{eqnarray}
Where $f(q)$ is the form factor which is related to the length of displacement.
If the waiting time distribution is exponential, $\psi(t)=\frac{1}{\tau}\exp (-t/\tau)$, where $\tau$ is the timescale of the 
distribution, then from above equation we get,
\begin{eqnarray}
 \phi^{single}_{CTRW}(q,s)&=&\frac{s\tau}{s(s\tau+1)}\frac{s\tau+1}{s\tau+1-f(q)}\nonumber\\
                    &=&\frac{1}{s+(1-f(q))/\tau}.
 \label{single_wt2-c}
\end{eqnarray}
Where $\hat \psi(s)=\frac{1}{1+s\tau}$. 
Eq.\ref{single_wt} in time plane can be written as,
\begin{equation}
 \phi^{single}_{CTRW}(q,t)=\exp (-t(1-f(q))/\tau).
\end{equation}

If we now consider that there are two different kinds of dynamics, like the activated
dynamics and the MCT like diffusive dynamics then we can assume that there are two different waiting time distributions, 
one is $\psi_1^0(t)$ and another is $\psi_2^0(t)$ \cite{ctrw_2dis}. 
The bare waiting time distributions get modified due to the presence of each other. The modified waiting time distributions can be 
written as,
\begin{eqnarray}
 \psi_1(t)=\psi^0_1(t)(1-\int_0^t \psi^0_2(t')dt')\nonumber\\
 \psi_2(t)= \psi^0_2(t)(1-\int_0^t \psi^0_1(t')dt').
\end{eqnarray}
According to CTRW formalism \cite{tachiya_bagchi_seki} we can write the structural relaxation as,
\begin{eqnarray}
 \phi^{double}_{CTRW}(q,s)&=&\frac{1-\hat\psi_1(s)-\hat \psi_2(s)}{s}\nonumber\\
                &\times&\Big(\frac{1}{1-f_1(q)\hat\psi_1(s)-f_2(q)\hat\psi_2(s)}\Big).
\label{two_wt-c}
\end{eqnarray}
If both the bare waiting time distributions are exponential {\it i.e} $\psi^0_{1,2}=\frac{1}{\tau_{1,2}}\exp(-t/\tau_{1,2})$ then  
the modified waiting time distributions become,
\begin{eqnarray}
 \psi_{1,2}(t)&=&\frac{1}{\tau_{1,2}}\exp(-t/\tau_{1,2})\Big[1-\int_0^t\frac{1}{\tau_{2,1}}\exp(-t'/\tau_{2,1})dt'\Big]\nonumber\\
              &=&\frac{1}{\tau_{1,2}}\exp(-(\frac{1}{\tau_1}+\frac{1}{\tau_2})t).     
\end{eqnarray}
In frequency plane they can be written as, 
\begin{eqnarray}
 \hat \psi_{1,2}(s)=\frac{\tau_{2,1}}{\tau_1\tau_2s+\tau_1+\tau_2}.
\end{eqnarray}
Therefore for both exponential waiting time distributions the structural relaxation $\phi^{double}_{CTRW}(q,s)$ becomes,
\begin{eqnarray}
 \phi^{double}_{CTRW}(q,s)&=&\frac{1-\frac{\tau_2}{\tau_1\tau_2s+\tau_1+\tau_2}-\frac{\tau_1}{\tau_1\tau_2s+\tau_1+\tau_2}}{s}\nonumber\\
             &\times&\frac{1}{1-\frac{f_1(q)\tau_2}{\tau_1\tau_2s+\tau_1+\tau_2}-\frac{f_2(q)\tau_1}{\tau_1\tau_2s+\tau_1+\tau_2}}\nonumber\\
             &=&\frac{\tau_1\tau_2s}{s(\tau_1\tau_2s+\tau_1+\tau_2)}\nonumber\\ 
             &\times&\frac{(\tau_1\tau_2s+\tau_1+\tau_2)}{\tau_1\tau_2s+\tau_1(1-f_2(q))+\tau_2(1-f_1(q)))}\nonumber\\
             &=&\frac{1}{s+(1-f_1(q))/\tau_1+(1-f_2(q))/\tau_2}.
\end{eqnarray}
In time plane the structural relaxation is the product: 
\begin{eqnarray}
 \phi^{double}_{CTRW}(q,t)&\simeq& \exp(-t(1-f_1(q))/\tau_1)\nonumber\\
         &\times& \exp(-t(1-f_2(q))/\tau_2) \nonumber\\
          &=& \phi_{hop}(q,t)\times \phi_{MCT}(q,t).
\label{ctrw_prod-c}
\end{eqnarray}
Note that this formalism in CTRW gives the same result as given by Eq.\ref{scheme-1} and thus similar to that given by Eq.\ref{scheme-2}.
Thus if we use this formalism then we will not get a SE breakdown.


\end{document}